\documentclass[preprint,showpacs]{revtex4}
\usepackage{graphicx}
\usepackage{bm}
\usepackage{color}
\usepackage{amsmath}
\usepackage{natbib}

\begin{document}

\title{Quantum kinetics derivation as generalization of the quantum hydrodynamics method}

\author{Pavel A. Andreev}
\email{andreevpa@physics.msu.ru}
\affiliation{M. V. Lomonosov Moscow State University, Moscow, Russia.}

 \date{\today}

\begin{abstract}

We present a new way of quantum kinetic equation derivation. This method appears as a natural generalization of the many-particle quantum hydrodynamic method. Kinetic equations are derived for different system of particles. First of all we consider quantum plasma and pay special attention to the spin evolution. We show that we need a set of two kinetic equations for description of spinning particles. One of these equations is the equation for distribution function, however this equation contains new function, even in the self-consistent field approximation. This is a spin-distribution function introduced in the paper. Therefore we have to derive kinetic equation for spin distribution function evolution, which is presented here and used to construct a closed set of kinetic equations. We also present kinetic equation for system of neutral particles with a short-range interaction in the first order by the interaction radius approximation. We derive a set of kinetic equations for particles having electric dipole moment, this set analogous to the equations set for spinning particles, but it has some differences. As a special topic we find kinetic equations for graphene carriers in the vicinity of the Dirac points. Derived equations, in general case, contain two-particle distribution functions, which take into account contribution of the quantum correlations including the exchange interaction, but we restrict ourself by the self-consistent field approximation to obtain closed kinetic description, in the system of particles with the short-range interaction.

 \end{abstract}

%  \pacs{52.35.We, 67.10.-j, 67.25.dk, 67.30.hj}% PACS, the Physics and Astronomy
                             % Classification Scheme.
% \keywords{}%Use showkeys class option if keyword

\maketitle

%%%%%%%%%%TEXT

\section{Introduction}

Quantum plasma studying requires developments of new theoretical
methods. In classical regime it has been very fruitful to use
hydrodynamical and kinetic methods. So, we can expect that their
quantum generalization will be as profound. In last decade the
method of quantum hydrodynamics has been developed by many teams
\cite{MaksimovTMP 1999}-\cite{Haas PRE 12}. The semi-relativistic
interactions, the spin-spin \cite{MaksimovTMP 2001}, the
spin-current \cite{Andreev RPJ 07}, the spin-orbit \cite{Andreev
Arxiv 333}, the current-current \cite{Ivanov arxiv 2012}
interactions, \emph{and} the Darwin term and the semi-relativistic
amendment to the kinetic energy \cite{Ivanov arxiv 2012}, have
been included in the quantum hydrodynamics scheme along with the
Coulomb interaction. Wigner kinetic approximation has been used
for studying of the quantum plasma \cite{Brodin PRL 08}-\cite{Haas
NJP 10}. A physical kinetics, free of any special assumptions on
the space-time geometry, was developed in Ref. \cite{Altaisky PL A
10}. A quantum distribution function such that calculating
statistical averages leads to the same local values of the number
of particles, the momentum, and the energy as those in quantum
mechanics was found in Ref. \cite{Maksimov TMP 2002}. This
distribution function coincides with the Wigner function only for
spatially homogeneous systems. It was used for derivation of an
quantum kinetic equation \cite{Maksimov TMP 2002}-\cite{Maksimov
TMP 05} and consideration of some physical problems \cite{Maximov
IJQC 04}-\cite{Maksimov P D 09}. Kinetic equation for charged
spinning particles defining the distribution function of ten
variables the coordinate $\textbf{r}$, the momentum $\textbf{p}$,
the spin $\textbf{s}$, and the time $t$ was considered in Ref.
\cite{Brodin PRL 08}. Integration of this distribution function
$f(\textbf{r},\textbf{p},\textbf{s},t)$ on the spin and momentum
gives concentration of particles $n(\textbf{r},t)$. Integrating
the product of the spin vector and the distribution function on
the same variables gives spin density. Quantum kinetic equation
for $f(\textbf{r},\textbf{p},\textbf{s},t)$ with the
semi-relativistic effects caused by interaction was suggested in
Ref. \cite{Asenjo arxiv 2011}. Obtaining of the kinetic equation
by averaging of the one-particle hydrodynamic equations on the
Maxwell or the Fermi distribution function was considered in Ref.
\cite{Kuzelev PU 99}. Set of equations consisting of the Vlasov
equation, including ponderomotive force acting of the magnetic
moments, for particle motion and Bargman-Michel-Telegdi (BMT)
equation, describing magnetic moment evolution, was used in Ref.
\cite{Vagin 09} to consider dispersion of elementary excitations
for the tensor gyromagnetic ratio. Review of the recent
achievements in the quantum kinetics can be found in Ref.s
\cite{P.K. Shukla UFN 10}, \cite{Shukla RMP 11}.

Microscopic density of particles in classical physics can be
presented as the sum of Dirac delta functions \cite{Klimontovich
book}-\cite{Kuz'menkov 91}
$$n(\textbf{r},t)=\sum_{n=1}^{N}\delta(\textbf{r}-\textbf{r}_{n}(t)),$$
where $N$ is the total number of particles in the system. This definition
of the particles concentration was used for construction of the quantum particles
concentration and derivation of the quantum hydrodynamic equations
\cite{MaksimovTMP 1999}, \cite{Andreev PRB 11}, \cite{Andreev
Arxiv 333}. Corresponding microscopic distribution function can be
written as
$$f=\sum_{n}\delta(\textbf{r}-\textbf{r}_{n}(t))\delta(\textbf{p}-\textbf{p}_{n}(t)).$$
Averaging of this function leads to macroscopic distribution
function allowing to derive a chain of relativistic kinetic
equations \cite{Kuz'menkov 91}. General form of quantum
distribution function might be presented in the following form
\cite{Maksimov TMP 2002}
$$\hat{f}(\textbf{r},\textbf{p})=\sum_{n} \hat{n}_{n}(\textbf{p})\hat{n}_{n}(\textbf{r}),$$
and it was expressed via the one-particle statistical operator.
However, we in this paper consider the quantum mechanical
averaging \cite{Landau Vol 3}
$$<L>=\int \psi^{*}\hat{L}\psi dR$$
of $\hat{f}$ defined as
\begin{equation}\label{QKin distr func operator intr} \hat{f}=\sum_{n}\delta(\textbf{r}-\widehat{\textbf{r}}_{n})\delta(\textbf{p}-\widehat{\textbf{p}}_{n}),\end{equation}
where $\widehat{\textbf{p}}_{n}$ is the momentum operator for
$n$-th particle, and $\textbf{p}$ is the numerical vector function
which arithmetizes the momentum space, as the coordinate
$\textbf{r}$ arithmetizes coordinate space.

We show that evolution of the quantum mechanical averaging of the
operator (\ref{QKin distr func operator intr}) leads to arising of
the second function, which is the quantum mechanical averaging of
the following operator
\begin{equation}\label{QKin distr spin func operator intr} \hat{S}^{\alpha}=\sum_{n}\widehat{\sigma}_{n}^{\alpha}\delta(\textbf{r}-\widehat{\textbf{r}}_{n})\delta(\textbf{p}-\widehat{\textbf{p}}_{n}).\end{equation}
Thus, we should use a set of two kinetic equations for description of spinning particles evolution.

We use this method for derivation of kinetic equations for
different physical systems. These are the spinning quantum plasma
briefly discussed above, charged or neutral particles having
electric dipole moment, neutral particles with the short-range
interaction, graphene carriers and graphene excitons.
Corresponding quantum hydrodynamic equation have been obtained
previously. At both the graphene carriers and the graphene
excitons description we have deal with the quasi-spin.
Consequently, we have to obtain a couple kinetic equations as for
spinning quantum plasma. At description of the electrically
polarized particles lead to appearing of a new function, analogous
to (\ref{QKin distr spin func operator intr}), but we should put
the electric dipole moment operator instead of the spin operator.

This paper is organized as follows.  In Sec. II  we start our
microscopic derivation of quantum kinetic equation. This
derivation is presented for the simples case of charged particles
in an external electric field described by scalar potential with
the Coulomb interaction. The self-consistent field approximation
considered for obtained kinetic equation. In Sec. III we introduce
the distribution function in a magnetic field and derive kinetic
equation for charged spinless particles in an external
electromagnetic field. In Sec. IV we consider kinetics of spinning
particles, we derive a couple of kinetic equations for the
distribution function and spin distribution function. In Sec. V we
obtain kinetic equations for particles having electric dipole
moment. In Sec. VI we study kinetic of neutral particles with the
short range interaction. In Sec. VII we find a set of kinetic
equation for carriers in the graphene. In Sec. VIII a set of kinetic
equations for the graphene excitons is obtained. In Sec. IX we
present the brief summary of our results. In Appendix we critically
examined some general method derivation of kinetic equation used as
for quantum and for classic systems.

\section{Construction of macroscopic equations}

\subsection{Kinetic equation for spinless particles: General form}

The equation of quantum kinetics is derived from the
non-stationary Schrodinger equation for system of N particles:
$$\imath\hbar\partial_{t}\psi(R,t)=\Biggl(\sum_{n}\biggl(\frac{1}{2m_{n}}\widehat{\textbf{D}}_{n}^{2}+e_{n}\varphi_{n,ext}\biggr)$$
\begin{equation}\label{QKin Hamiltonian}+\frac{1}{2}\sum_{n,k\neq
k}e_{n}e_{k}G_{nk}\Biggr)\psi(R,t).\end{equation} The following
designations are used in the equation (\ref{QKin Hamiltonian}):
$D_{n}^{\alpha}=-\imath\hbar\partial_{n}^{\alpha}-e_{n}A_{n,ext}^{\alpha}/c$,
$\varphi_{n,ext}$, $A_{n,ext}^{\alpha}$ are the potentials of the
external electromagnetic field,
$\partial_{n}^{\alpha}=\nabla_{n}^{\alpha}$ is the derivative on
the space variables of $n$-th particle, and $G_{nk}=1/r_{nk}$  is
the Green functions of the Coulomb interaction, $\psi(R,t)$ is the
psi function of N particle system,
$R=(\textbf{r}_{1},...,\textbf{r}_{N})$, $e_{n}$, $m_{n}$ are the
charge and the mass of particle, $\hbar$ is the Planck constant
and $c$ is the speed of light.

The first step in derivation of the kinetic equation is a definition of distribution function of particles.
We determine the distribution function of particles as the quantum-mechanical
average of the operator of distribution function:
$$\hat{f}=\sum_{n}\delta(\textbf{r}-\textbf{r}_{n})\delta(\textbf{p}-\widehat{\textbf{p}}_{n}).$$
This function is the microscopic distribution function in
classic physics, $\delta(\textbf{r})$-is the Dirac's
$\delta$-function.

In that way, the distribution function has form:
\begin{equation}\label{QKin def distribution function non sym}f(\textbf{r}, \textbf{p},t)=\frac{1}{2}\int \psi^{*}(R,t)\sum_{n}\biggl(\delta(\textbf{r}-\textbf{r}_{n})\delta(\textbf{p}-\widehat{\textbf{p}}_{n})+\delta(\textbf{p}-\widehat{\textbf{p}}_{n})\delta(\textbf{r}-\textbf{r}_{n})\biggr)\psi(R,t)dR,\end{equation}
where $dR=\prod_{n=1}^{N}d\textbf{r}_{n}$. This definition symmetric relatively to operators $\delta(\textbf{r}-\textbf{r}_{n})$ and $\delta(\textbf{p}-\widehat{\textbf{p}}_{n})$, but it still is not fully symmetric. To get the fully symmetric definition of the distribution function we need to add complex conjugated quantity. Thus we have following definition
\begin{equation}\label{QKin def distribution function}f(\textbf{r}, \textbf{p},t)=\frac{1}{4}\int \Biggl(\psi^{*}(R,t)\sum_{n}\biggl(\delta(\textbf{r}-\textbf{r}_{n})\delta(\textbf{p}-\widehat{\textbf{p}}_{n})+\delta(\textbf{p}-\widehat{\textbf{p}}_{n})\delta(\textbf{r}-\textbf{r}_{n})\biggr)\psi(R,t)+c.c.\Biggr)dR,\end{equation}
where c.c. stands for the complex conjugation.

For the first step we will consider set of the charged particles in external quasi-static electric field including the Coulomb interaction between particles, this means that we do not include the vector potential. Next, we will consider electromagnetic external field, and we will also consider spinning particles with spin-spin interaction.

Integrating of the distribution function over momentum we have concentration
\begin{equation}\label{QKin concentration}n(\textbf{r},t)=\int f(\textbf{r}, \textbf{p},t) d\textbf{p}\end{equation}
where the concentration has familiar from the quantum hydrodynamics form \cite{Andreev PRB 11}, \cite{MaksimovTMP 1999}
\begin{equation}\label{NBS def density}n(\textbf{r},t)=\int dR\sum_{n}\delta(\textbf{r}-\textbf{r}_{n})\psi^{*}(R,t)\psi(R,t),\end{equation}

Integrating of the product of momentum on the distribution function over momentum we find particles current
\begin{equation}\label{QKin current}\textbf{j}(\textbf{r},t)=\int \textbf{p}f(\textbf{r}, \textbf{p},t) d\textbf{p},\end{equation}
details of calculations you can find in appendix.

In the same way we have the kinetic energy
\begin{equation}\label{QKin kin energy}\varepsilon_{kin}(\textbf{r},t)=\int \frac{\textbf{p}^{2}}{2m}f(\textbf{r}, \textbf{p},t) d\textbf{p}.\end{equation}

In the absence of the inter-particle interaction we find
\begin{equation}\label{QKin kinetic equation gen without A}\partial_{t}f+\frac{1}{m}\textbf{p}\partial_{\textbf{r}}f+e\frac{\imath}{\hbar}\varphi(\textbf{r},t)\sin(\overleftarrow{\nabla}_{\textbf{r}}\nabla_{\textbf{p}})f=0,\end{equation}
where
$$\sin(\overleftarrow{\nabla}_{\textbf{r}}\nabla_{\textbf{p}})=\sum_{l=1}^{\infty}\frac{(\imath\hbar)^{2l+1}}{(2l+1)!}(\overleftarrow{\nabla}_{\textbf{r}}\nabla_{\textbf{p}})^{2l+1}.$$
In equation (\ref{QKin kinetic equation gen without A}) we have
used designation
$\varphi(\textbf{r},t)\overleftarrow{\nabla}_{\textbf{r}}$, it
means that the gradient operator$\nabla$ on spatial variables acts
on the left-hand side, instead of usual acting of operators on
function standing on the right-hand side. We do not write the
Plank constant $\hbar$ in the argument of $\sin$ to make this
notation more handy.

The last term appears here due to the commutation of the electric scalar potential with the $\delta(\textbf{p}-\widehat{\textbf{p}}_{n})$. At $l\geq 2$ terms are proportional to the plank constant $\sim\hbar^{l-1}$. They disappear in the classical limit $\hbar\rightarrow 0$.

In the quasi-classical limit we have to include one term of the sum in the last term only, so we have
\begin{equation}\label{QKin kinetic equation gen without A classic limit}\partial_{t}f+\frac{1}{m}\textbf{p}\partial_{\textbf{r}}f+e\textbf{E}\nabla_{\textbf{p}}f=0,\end{equation}
where we have written $\textbf{E}=-\partial_{\textbf{r}}\varphi$.

We also admit that distribution function (\ref{QKin def
distribution function non sym}) satisfy to the following equation
$$\partial_{t}f+\frac{1}{m}\textbf{p}\partial_{\textbf{r}}f+e\frac{\imath}{\hbar}\varphi(\textbf{r},t) \sum_{l=1}^{\infty}\frac{(\imath\hbar)^{l}}{l!}(\overleftarrow{\nabla}_{\textbf{r}}\nabla_{\textbf{p}})^{l}f=0.$$

At derivation of kinetic equation the distribution current arises in the second terms of equations (\ref{QKin kinetic equation gen without A}) and (\ref{QKin kinetic equation gen without A classic limit})
$$J^{\alpha}(\textbf{r}, \textbf{p},t)=\frac{1}{8}\int \Biggl((\hat{p}^{\alpha}_{n}\psi^{*}(R,t))\sum_{n}\biggl(\delta(\textbf{r}-\textbf{r}_{n})\delta(\textbf{p}-\widehat{\textbf{p}}_{n})+\delta(\textbf{p}-\widehat{\textbf{p}}_{n})\delta(\textbf{r}-\textbf{r}_{n})\biggr)\psi(R,t)$$
\begin{equation}\label{QKin def kinetic current}+\psi^{*}(R,t)\sum_{n}\biggl(\delta(\textbf{r}-\textbf{r}_{n})\delta(\textbf{p}-\widehat{\textbf{p}}_{n})+\delta(\textbf{p}-\widehat{\textbf{p}}_{n})\delta(\textbf{r}-\textbf{r}_{n})\biggr)\hat{p}^{\alpha}_{n}\psi(R,t)+c.c.\Biggr)dR,\end{equation}
which we approximately present in the following
\begin{equation}\label{QKin def kinetic current short}J^{\alpha}(\textbf{r}, \textbf{p},t)=p^{\alpha}f(\textbf{r}, \textbf{p},t),\end{equation}
which corresponds to the traditional structure of kinetic equation.
The last formula has been already used in formulas (\ref{QKin kinetic equation gen without A}) and (\ref{QKin kinetic equation gen without A classic limit}).

Integrating equation (\ref{QKin kinetic equation gen without A}) and other form of kinetic equation obtained in this paper we can find that they give us the equations of the many-particle quantum hydrodynamics obtained in Ref.s  \cite{MaksimovTMP 1999}, \cite{Andreev PRB 11}, \cite{Andreev Arxiv 333}.

Now we are going to include the Coulomb inter-particle interaction
$$\partial_{t}f+\frac{1}{m}\textbf{p}\partial_{\textbf{r}}f+e\frac{\imath}{\hbar}\varphi\sin(\overleftarrow{\nabla}_{\textbf{r}}\nabla_{\textbf{p}})f$$
$$+\frac{1}{4}\frac{\imath}{\hbar}
\int \sum_{n,k\neq n}(e_{n}e_{k}G_{nk})\sin(\overleftarrow{\nabla}_{\textbf{r}, n}\nabla_{\textbf{p}})\times$$
\begin{equation}\label{QKin kinetic equation gen with cul int}\times\Biggl(\psi^{*}(R,t)\biggl(\delta(\textbf{r}-\textbf{r}_{n})\delta(\textbf{p}-\widehat{\textbf{p}}_{n})+\delta(\textbf{p}-\widehat{\textbf{p}}_{n})\delta(\textbf{r}-\textbf{r}_{n})\biggr)\psi(R,t)+c.c.\Biggr)dR
=0.\end{equation}
In equation (\ref{QKin kinetic equation gen with cul int}) we find one more term in comparison with equation (\ref{QKin kinetic equation gen without A}). This new term is the last term in equation (\ref{QKin kinetic equation gen with cul int}), it is caused by the Coulomb interaction.

Equation (\ref{QKin kinetic equation gen with cul int}) can be rewritten using two-particle function
$$\partial_{t}f+\frac{1}{m}\textbf{p}\partial_{\textbf{r}}f+e\frac{\imath}{\hbar}\varphi\sin(\overleftarrow{\nabla}_{\textbf{r}}\nabla_{\textbf{p}})f$$
\begin{equation}\label{QKin kinetic equation gen with cul int and two part f}+e^{2}\frac{\imath}{\hbar}
\int G(\textbf{r}-\textbf{r}')\sin(\overleftarrow{\nabla_{\textbf{r}}}\nabla_{\textbf{p}}) f_{2}(\textbf{r},\textbf{p},\textbf{r}',\textbf{p}',t) d\textbf{r}'d\textbf{p}'
=0,\end{equation}
where
$$f_{2}(\textbf{r},\textbf{p},\textbf{r}',\textbf{p}',t)=\frac{1}{4}\int \Biggl(\psi^{*}(R,t)\sum_{n,k\neq n}\delta(\textbf{r}'-\textbf{r}_{k})\delta(\textbf{p}'-\widehat{\textbf{p}}_{k})\times$$
\begin{equation}\label{QKin two part distr func}\times\biggl(\delta(\textbf{r}-\textbf{r}_{n})\delta(\textbf{p}-\widehat{\textbf{p}}_{n})+\delta(\textbf{p}-\widehat{\textbf{p}}_{n})\delta(\textbf{r}-\textbf{r}_{n})\biggr)\psi(R,t)+c.c.\Biggr)dR,\end{equation}
is the two-particle distribution function.

In the quasi-classical limit we have to include one term of the sum in the last term only, so we have
\begin{equation}\label{QKin kinetic equation gen without A with int classic limit}\partial_{t}f+\frac{1}{m}\textbf{p}\partial_{\textbf{r}}f+e\textbf{E}\nabla_{\textbf{p}}f-e^{2}
\int
\nabla_{\textbf{r}}G(\textbf{r}-\textbf{r}')\cdot\nabla_{\textbf{p}}
f_{2}(\textbf{r},\textbf{p},\textbf{r}',\textbf{p}',t)
d\textbf{r}'d\textbf{p}' =0.\end{equation}
Equation (\ref{QKin
kinetic equation gen without A with int classic limit}) looks like
classical kinetic equation. However we should admit that
two-particle distribution function $f_{2}$ is defined via wave
function $\psi$. Therefore, we see, analogously to the quantum
hydrodynamics \cite{MaksimovTMP 1999}, \cite{Andreev PRB 11}, this
equation is a quantum kinetic equation containing information
about quantum effects, in particular, the exchange interaction.
Some methods of obtaining of the closed quantum kinetic description was discussed and developed in Ref. \cite{Petrus P A 76}.

%\textbf{quasi-classical limit}

\subsection{The self-consistent field approximation}

Considering many-particle quantum kinetics we find that equation of distribution function $f$ evolution contains two-particle distribution function $f_{2}$ that corresponds to the classical kinetics. For particles with a long-range interaction, in our case it is the Coulomb interaction, we can introduce, well-known in classical physics and quantum hydrodynamics, the self-consistent field approximation. In this approximation we separate two-particle distribution function $f_{2}$ in the product of two one-particle distribution functions $f$
\begin{equation}\label{QKin two part dist func-selfC F appr}f_{2}(\textbf{r},\textbf{p},\textbf{r}',\textbf{p}',t)=f(\textbf{r}, \textbf{p},t) f(\textbf{r}', \textbf{p}',t).\end{equation}
This approximation gives us closed mathematical apparatus in the form of one nonlinear integral equation, which
in the quasi-classic limit to be
\begin{equation}\label{QKin kinetic equation gen without A classic limit}\partial_{t}f+\frac{1}{m}\textbf{p}\partial_{\textbf{r}}f+e\textbf{E}\nabla_{\textbf{p}}f-e^{2}
\int \nabla_{\textbf{r}}G(\textbf{r}-\textbf{r}') f(\textbf{r}',\textbf{p}',t) d\textbf{r}'d\textbf{p}' \cdot\nabla_{\textbf{p}}f(\textbf{r},\textbf{p},t)
=0.\end{equation}

In general case, from equation (\ref{QKin kinetic equation gen with cul int and two part f}), in the self-consistent field approximation, we find
$$\partial_{t}f+\frac{1}{m}\textbf{p}\partial_{\textbf{r}}f+e\frac{\imath}{\hbar}\varphi\sin(\overleftarrow{\nabla}_{\textbf{r}}\nabla_{\textbf{p}})f$$
\begin{equation}\label{QKin kinetic equation gen with cul int and two part f}+e^{2}\frac{\imath}{\hbar}
\int G(\textbf{r}-\textbf{r}')\sin(\overleftarrow{\nabla_{\textbf{r}}}\nabla_{\textbf{p}})f(\textbf{r},\textbf{p},t) f(\textbf{r}',\textbf{p}',t) d\textbf{r}'d\textbf{p}'
=0,\end{equation}

Introducing electric field for inter-particle interaction we come to the set of kinetic equation and field equations (Maxwell equations) instead of integro-differential equation (\ref{QKin kinetic equation gen without A classic limit}). This set appears as
\begin{equation}\label{QKin kinetic equation gen without A classic limit with E}\partial_{t}f+\frac{1}{m}\textbf{p}\partial_{\textbf{r}}f+e\textbf{E}\nabla_{\textbf{p}}f=0,\end{equation}
\begin{equation}\label{QKin electro stat Max} \begin{array}{ccc}\nabla\times \textbf{E}=0,&   \nabla\textbf{E}=4\pi e\int f(\textbf{r},\textbf{p},t)d\textbf{p}.\end{array}\end{equation}
In the result we see that derived equations corresponds to the
Vlasov equation. It also coincides with the results obtained in
Ref.s \cite{Marklund TTSP 11}-\cite{Zamanian NJP 10}, where was
used the Wigner distribution function.

\section{Charged spinless particles in magnetic field}

Studying charged particles in an magnetic field we have long derivative $\textbf{D}_{n}=-\imath\hbar\nabla_{n}-e\textbf{A}_{n}/c$ instead of the short one $-\imath\hbar\nabla_{n}$. So, we have to include it at definition of the distribution function $f(\textbf{r},\textbf{p},t)$ and in this case we have
\begin{equation}\label{QKin def distribution function in magn field}f(\textbf{r}, \textbf{p},t)=\frac{1}{4}\int \Biggl(\psi^{*}(R,t)\sum_{n}\biggl(\delta(\textbf{r}-\textbf{r}_{n})\delta(\textbf{p}-\widehat{\textbf{D}}_{n})+\delta(\textbf{p}-\widehat{\textbf{D}}_{n})\delta(\textbf{r}-\textbf{r}_{n})\biggr)\psi(R,t)+c.c.\Biggr)dR,\end{equation}

In the absence of the inter-particle interaction we find
$$\partial_{t}f+\frac{1}{m}\textbf{p}\partial_{\textbf{r}}f-\frac{e}{mc}\partial^{\alpha}A^{\beta}\partial^{\beta}_{\textbf{p}}\biggl(p^{\alpha}f\biggr)-\frac{e}{c}(\partial_{t}\textbf{A})\nabla_{\textbf{p}}f$$
\begin{equation}\label{QKin kinetic equation gen with A}-\frac{e}{mc}\frac{\imath}{\hbar}A^{\alpha}\sin(\overleftarrow{\nabla}_{\textbf{r}}\nabla_{\textbf{p}})(p^{\alpha}f)+e\frac{\imath}{\hbar}\varphi\sin(\overleftarrow{\nabla}_{\textbf{r}}\nabla_{\textbf{p}})f=0 .\end{equation}
Including in consideration of the vector potential of the external electromagnetic field leads to appearing of three additional terms in the kinetic equation (\ref{QKin kinetic equation gen with A}), which are the third-fifth terms. The fourth term gives contribution in the force acting on the charge of the external electric field. The third and fifth terms is the magnetic part of the quantum Lorentz force. In the quasi-classical limit it has well-known form presented below.

In the quasi-classical limit we have to include one term of the sum in the last term only, so we have
\begin{equation}\label{QKin kinetic equation gen with A classic limit}\partial_{t}f+\frac{\textbf{p}}{m}\partial_{\textbf{r}}f+e\biggl(\textbf{E}+\frac{1}{mc}\textbf{p}\times\textbf{B}\biggr)\nabla_{\textbf{p}}f=0,\end{equation}
where we have written $\textbf{E}=-\partial_{\textbf{r}}\varphi-\partial_{t}\textbf{A}/c$.

We should write the Lorentz force in equation (\ref{QKin kinetic
equation gen with A classic limit}) via velocity
$\textbf{p}/m=\textbf{v}$, but distribution function
$f=f(\textbf{r},\textbf{p},t)$ depends on momentum, and we write
all coefficients via momentum $\textbf{p}$.

\section{Kinetic evolution of spinning particles}

In section III we present the Schrodinger equation contained only
the Coulomb interaction. Now we going to consider kinetics of
spinning charged particles. In this section we are particularly
interested in the spin-spin interactions. The whole Hamiltonian,
used in this case reads
\begin{widetext}
$$\hat{H}=\sum_{n}\biggl(\frac{1}{2m_{n}}\hat{D}^{\alpha}_{n}\hat{D}^{\alpha}_{n}+e_{n}\varphi^{ext}_{n}-\gamma_{n}\widehat{\sigma}^{\alpha}_{n}B^{\alpha}_{n(ext)}\biggr)$$
\begin{equation}\label{QKin ham gen}+\frac{1}{2}\sum_{k,n\neq p}(e_{k}e_{n}G_{kn}-\gamma_{k}\gamma_{n}G^{\alpha\beta}_{kn}\widehat{\sigma}^{\alpha}_{k}\widehat{\sigma}^{\beta}_{n}),\end{equation}
where
$$\hat{D}_{n}^{\alpha}=-\imath\hbar\partial_{n}^{\alpha}-\frac{e_{n}}{c}A_{n,ext}^{\alpha}.$$
\end{widetext}
In these formulas we have used following designations. The Green's
functions of the Coulomb, the spin-spin and the spin-current
interactions has the following form $G_{pn}=1/r_{pn}$,
$G^{\alpha\beta}_{pn}=4\pi\delta^{\alpha\beta}\delta(\textbf{r}_{pn})+\partial^{\alpha}_{p}\partial^{\beta}_{p}(1/r_{pn})$,
where $\gamma_{p}$  is the gyromagnetic ratio. For electrons
$\gamma_{p}$ reads $\gamma_{p}=e_{p}\hbar/(2m_{p}c)$,
$e_{p}=-|e|$. The quantities
$\varphi^{ext}_{p}=\varphi(\textbf{r}_{p},t),A^{\alpha}_{p
(ext)}=A^{\alpha}(\textbf{r}_{p},t)$ are the scalar and the vector
potentials of the external electromagnetic field:
$$ B^{\alpha}_{
(ext)}(\textbf{r}_{p},t)=\varepsilon^{\alpha\beta\gamma}\nabla^{\beta}_{p}A^{\gamma}_{(ext)}(\textbf{r}_{p},t),
$$
$$E^{\alpha}_{(ext)}(\textbf{r}_{p},t)=-\nabla^{\alpha}_{p}\varphi_{ext}(\textbf{r}_{p},t)-
 \frac{1}{c}\partial_{t}A^{\alpha}_{ext}(\textbf{r}_{p},t) .$$
$\widehat{\sigma}^{\alpha}_{p}$ is the Pauli matrix, a commutation
relations for them is
$$[\widehat{\sigma}^{\alpha}_{p},\widehat{\sigma}^{\beta}_{n}]=2\imath\delta_{pn}\varepsilon^{\alpha\beta\gamma}\widehat{\sigma}^{\gamma}_{p}.$$

First we present kinetic equation for spinning particles in the absence of the inter-particle interaction
$$\partial_{t}f+\frac{1}{m}\textbf{p}\partial_{\textbf{r}}f-\frac{e}{mc}\partial^{\alpha}A^{\beta}\partial^{\beta}_{\textbf{p}}\biggl(p^{\alpha}f\biggr)-\frac{e}{c}(\partial_{t}\textbf{A})\nabla_{\textbf{p}}f$$
$$-\frac{e}{mc}\frac{\imath}{\hbar}A^{\alpha}\sin(\overleftarrow{\nabla}_{\textbf{r}}\nabla_{\textbf{p}})(p^{\alpha}f)+e\frac{\imath}{\hbar}\varphi\sin(\overleftarrow{\nabla}_{\textbf{r}}\nabla_{\textbf{p}})f$$
\begin{equation}\label{QKin kinetic equation gen with spin}-\mu\frac{\imath}{\hbar} B^{\alpha}\sin(\overleftarrow{\nabla}_{\textbf{r}}\nabla_{\textbf{p}})S^{\alpha}(\textbf{r}, \textbf{p},t)=0\end{equation}
In the quasi-classical limit we have to include one term of the
sum in the last term only, so we have
\begin{equation}\label{QKin kinetic equation gen with spin semi classic limit}\partial_{t}f+\frac{\textbf{p}}{m}\partial_{\textbf{r}}f+e\biggl(\textbf{E}+\frac{1}{mc}\textbf{p}\times\textbf{B}\biggr)\nabla_{\textbf{p}}f+\partial_{\alpha} B^{\beta}(\textbf{r},t)\partial_{\textbf{p}\alpha} S^{\beta}(\textbf{r}, \textbf{p},t)=0,\end{equation}
where we have new quantity in these equations, we can call it spin-distribution function, it's explicit form is
\begin{equation}\label{QKin def spin distribution function}S^{\alpha}(\textbf{r}, \textbf{p},t)=\frac{1}{4}\int \Biggl(\psi^{*}(R,t)\sum_{n}\biggl(\delta(\textbf{r}-\textbf{r}_{n})\delta(\textbf{p}-\widehat{\textbf{D}}_{n})+\delta(\textbf{p}-\widehat{\textbf{D}}_{n})\delta(\textbf{r}-\textbf{r}_{n})\biggr)\sigma^{\alpha}_{n}\psi(R,t)+h.c.\Biggr)dR,\end{equation}
where h.c. stands for the Hermitian conjugation.
$S^{\alpha}(\textbf{r}, \textbf{p},t)$ is kinetic analog of the spin density, which arises in the quantum hydrodynamics \cite{MaksimovTMP 2001}, \cite{Andreev RPJ 07} and has form
\begin{equation}\label{QKin def spin density}S^{\alpha}(\textbf{r},t)=\int dR\sum_{n}\delta(\textbf{r}-\textbf{r}_{n})\psi^{*}(R,t)\widehat{\sigma}^{\alpha}_{n}\psi(R,t),\end{equation}
we have used same letter for designation of the spin density and the spin-distribution function, but they differ by set of arguments.

It can be shown that
\begin{equation}\label{QKin def spin density connection of S and S} S^{\alpha}(\textbf{r}, t)=\int S^{\alpha}(\textbf{r}, \textbf{p},t)d\textbf{p}.\end{equation}

Magnetization $M^{\alpha}(\textbf{r},t)$ usually used in the
quantum hydrodynamics \cite{MaksimovTMP 2001}, \cite{Andreev RPJ
07}, \cite{Andreev Arxiv 333}, and \cite{Andreev arxiv 12
SpinCurrent}. Magnetization $M^{\alpha}(\textbf{r},t)$ has simple
connection with the spin density $M^{\alpha}(\textbf{r},t)=\gamma
S^{\alpha}(\textbf{r},t)$, where $\gamma$ is the gyromagnetic
ratio for considering species of particles.

To get complete description of  systems of spinning particles we
have to derive an equation for the spin-distribution function
$S^{\alpha}(\textbf{r}, \textbf{p},t)$.

\subsection{Including of interaction}

Introducing of interaction makes  the kinetic equation more larger, but we present it's explicit form
$$\partial_{t}f+\frac{1}{m}\textbf{p}\partial_{\textbf{r}}f-\frac{e}{mc}\partial^{\alpha}A^{\beta}\partial^{\beta}_{\textbf{p}}\biggl(p^{\alpha}f\biggr)-\frac{e}{c}(\partial_{t}\textbf{A})\nabla_{\textbf{p}}f$$
$$-\frac{e}{mc}\frac{\imath}{\hbar}A^{\alpha}\sin(\overleftarrow{\nabla}_{\textbf{r}}\nabla_{\textbf{p}})(p^{\alpha}f)+e\frac{\imath}{\hbar}\varphi\sin(\overleftarrow{\nabla}_{\textbf{r}}\nabla_{\textbf{p}})f$$
$$-\mu\frac{\imath}{\hbar} B^{\alpha}\sin(\overleftarrow{\nabla}_{\textbf{r}}\nabla_{\textbf{p}})S^{\alpha}(\textbf{r}, \textbf{p},t)$$
$$+\frac{1}{4}\frac{\imath}{\hbar}
\int \sum_{n,k\neq n}(e_{n}e_{k}G_{nk})\sin(\overleftarrow{\nabla}_{\textbf{r}n}\nabla_{\textbf{p}})\times$$
$$\times\Biggl(\psi^{*}(R,t)\biggl(\delta(\textbf{r}-\textbf{r}_{n})\delta(\textbf{p}-\widehat{\textbf{D}}_{n})+\delta(\textbf{p}-\widehat{\textbf{D}}_{n})\delta(\textbf{r}-\textbf{r}_{n})\biggr)\psi(R,t)+h.c.\Biggr)dR
$$
$$+\frac{1}{4}\frac{\imath}{\hbar}
\int \sum_{n,k\neq n}(\gamma_{n}\gamma_{k}G^{\mu\nu}_{nk})\sin(\overleftarrow{\nabla}_{\textbf{r}n}\nabla_{\textbf{p}})\times$$
\begin{equation}\label{QKin kinetic equation gen with spin and int}\times\Biggl(\psi^{*}(R,t)\biggl(\delta(\textbf{r}-\textbf{r}_{n})\delta(\textbf{p}-\widehat{\textbf{D}}_{n})+\delta(\textbf{p}-\widehat{\textbf{D}}_{n})\delta(\textbf{r}-\textbf{r}_{n})\biggr)\sigma_{n}^{\mu}\sigma_{k}^{\nu}\psi(R,t)+h.c.\Biggr)dR
=0.\end{equation} Let's describe meaning of terms in this
equation. The first seven terms coincide with the same terms in
the kinetic equation for charged spinless particles (\ref{QKin
kinetic equation gen with A}). They are time evolution of
distribution function presented by the  first term, convective
part of distribution function evolution,  presented by the second
and third terms, correspondingly. Terms four, and six give
quantum-kinetic generalization of the Lorentz force describing
action of the external magnetic field on particle charges. The
fifth and seventh terms show the force acting on charges exerted
by the external electric field. The eighth term describes action
of the external magnetic field on spin of particles entering
equation via the spin-distribution function
$S^{\alpha}(\textbf{r},\textbf{p},t)$. The last two terms present
inter-particle interaction, the Coulomb and the spin-spin
interactions correspondingly, and it's influence on evolution of
the distribution function.

Introducing two-particle spin-distribution function
as
$$S_{2}^{\alpha\beta}(\textbf{r},\textbf{p},\textbf{r}',\textbf{p}',t)=\frac{1}{4}\int \Biggl(\psi^{*}(R,t)\sum_{n,k\neq n}\delta(\textbf{r}'-\textbf{r}_{k})\delta(\textbf{p}'-\widehat{\textbf{D}}_{k})\times$$
\begin{equation}\label{QKin two part distr func}\times\biggl(\delta(\textbf{r}-\textbf{r}_{n})\delta(\textbf{p}-\widehat{\textbf{D}}_{n})+\delta(\textbf{p}-\widehat{\textbf{D}}_{n})\delta(\textbf{r}-\textbf{r}_{n})\biggr)\sigma_{n}^{\alpha}\sigma_{k}^{\beta}\psi(R,t)+h.c.\Biggr)dR\end{equation}
we can rewrite previous equation in the following form
$$\partial_{t}f+\frac{1}{m}\textbf{p}\partial_{\textbf{r}}f-\frac{e}{mc}\partial^{\alpha}A^{\beta}\partial^{\beta}_{\textbf{p}}\biggl(p^{\alpha}f\biggr)-\frac{e}{c}(\partial_{t}\textbf{A})\nabla_{\textbf{p}}f$$
$$-\frac{e}{mc}\frac{\imath}{\hbar}A^{\alpha}\sin(\overleftarrow{\nabla}_{\textbf{r}}\nabla_{\textbf{p}})(p^{\alpha}f)
+e\frac{\imath}{\hbar}\varphi\sin(\overleftarrow{\nabla}_{\textbf{r}}\nabla_{\textbf{p}})f$$
$$-\mu\frac{\imath}{\hbar} B^{\alpha}\sin(\overleftarrow{\nabla}_{\textbf{r}}\nabla_{\textbf{p}})S^{\alpha}(\textbf{r}, \textbf{p},t)$$
$$+e^{2}\frac{\imath}{\hbar}
\int G(\textbf{r}-\textbf{r}')\sin(\overleftarrow{\nabla_{\textbf{r}}}\nabla_{\textbf{p}}) f_{2}(\textbf{r},\textbf{p},\textbf{r}',\textbf{p}',t) d\textbf{r}'d\textbf{p}'$$
\begin{equation}\label{QKin kinetic equation gen with spin and int with two part F}+\gamma^{2}\frac{\imath}{\hbar}
\int G^{\mu\nu}(\textbf{r}-\textbf{r}')\sin(\overleftarrow{\nabla_{\textbf{r}}}\nabla_{\textbf{p}}) S^{\mu\nu}_{2}(\textbf{r},\textbf{p},\textbf{r}',\textbf{p}',t) d\textbf{r}'d\textbf{p}'=0\end{equation}

Using of the two-particle functions we have the two last terms in equation (\ref{QKin kinetic equation gen with spin and int with two part F}) written shortly. It also gives a form of kinetic equation analogous to the BBGKY equations, that allows us use well-known and useful ideas to get closed set of equations, such as introduction of the self-consistent field approximation, which we discuss for spinning below.

\subsection{spin-distribution function evolution}

In kinetic equation for spinning particles appears the spin distribution function. Therefore, for construction of the closed set of equation describing spinning particles we have to find equation evolution of the spin distribution function. For this goal we differentiate spin distribution function with respect to time, after some calculations we find the kinetic equation for spin distribution function evolution
$$\partial_{t}S^{\alpha}(\textbf{r},\textbf{p},t)+\frac{1}{m}\textbf{p}\partial_{\textbf{r}}S^{\alpha}-\frac{e}{mc}\partial^{\gamma}A^{\beta}\partial^{\beta}_{\textbf{p}}\biggl(p^{\gamma}S^{\alpha}\biggr)-\frac{e}{c}(\partial_{t}\textbf{A})\nabla_{\textbf{p}}S^{\alpha}$$
$$-\frac{e}{mc}\frac{\imath}{\hbar}A^{\beta}\sin(\overleftarrow{\nabla}_{\textbf{r}}\nabla_{\textbf{p}})(p^{\beta}S^{\alpha})
+e\frac{\imath}{\hbar}\varphi\sin(\overleftarrow{\nabla}\nabla_{\textbf{p}})S^{\alpha}$$
$$-\mu\frac{\imath}{\hbar} B^{\alpha}\sin(\overleftarrow{\nabla}\nabla_{\textbf{p}}) f(\textbf{r}, \textbf{p},t)$$
$$+e^{2}\frac{\imath}{\hbar}
\int G(\textbf{r}-\textbf{r}')\sin(\overleftarrow{\nabla_{\textbf{r}}}\nabla_{\textbf{p}}) M_{2}(\textbf{r},\textbf{p},\textbf{r}',\textbf{p}',t) d\textbf{r}'d\textbf{p}'$$
$$+\gamma^{2}\frac{\imath}{\hbar}
\int G^{\alpha\beta}(\textbf{r}-\textbf{r}')\sin(\overleftarrow{\nabla_{\textbf{r}}}\nabla_{\textbf{p}}) N^{\beta}_{2}(\textbf{r},\textbf{p},\textbf{r}',\textbf{p}',t) d\textbf{r}'d\textbf{p}'$$
\begin{equation}\label{QKin kinetic equation (spin evol) gen with spin and int with two part F}-\frac{2\gamma}{\hbar}\varepsilon^{\alpha\beta\gamma}\Biggl(B^{\beta}S^{\gamma}+\gamma\int G^{\beta\delta}(\textbf{r},\textbf{r}')S_{2}^{\gamma\delta}(\textbf{r},\textbf{p},\textbf{r}',\textbf{p}',t)d\textbf{r}'d\textbf{p}'\Biggr).\end{equation}
There is no high order derivatives on space variables in the last term. In the second term of equation (\ref{QKin kinetic equation (spin evol) gen with spin and int with two part F}) we have used following presentation for the kinetic spin current
\begin{equation}\label{QKin kin spin curr def}J^{\alpha\beta}(\textbf{r},\textbf{p},t)=p^{\beta}S^{\alpha}(\textbf{r},\textbf{p},t),\end{equation}
which has following explicit form
$$J^{\alpha\beta}(\textbf{r}, \textbf{p},t)=\frac{1}{8}\int \Biggl((\hat{D}^{\beta}_{n}\psi^{*}(R,t))\sum_{n}\biggl(\delta(\textbf{r}-\textbf{r}_{n})\delta(\textbf{p}-\widehat{\textbf{D}}_{n})+\delta(\textbf{p}-\widehat{\textbf{D}}_{n})\delta(\textbf{r}-\textbf{r}_{n})\biggr)\widehat{\sigma}_{n}^{\alpha}\psi(R,t)$$
\begin{equation}\label{QKin kin spin curr def}+\psi^{*}(R,t)\sum_{n}\biggl(\delta(\textbf{r}-\textbf{r}_{n})\delta(\textbf{p}-\widehat{\textbf{D}}_{n})+\delta(\textbf{p}-\widehat{\textbf{D}}_{n})\delta(\textbf{r}-\textbf{r}_{n})\biggr)\hat{D}^{\beta}\widehat{\sigma}_{n}^{\alpha}\psi(R,t)+h.c.\Biggr)dR.\end{equation}

We have used a couple of new two-particle functions
$$M_{2}^{\alpha}(\textbf{r},\textbf{p},\textbf{r}',\textbf{p}',t)=\frac{1}{4}\int \Biggl(\psi^{*}(R,t)\sum_{n,k\neq n}\delta(\textbf{r}'-\textbf{r}_{k})\delta(\textbf{p}'-\widehat{\textbf{D}}_{k})\times$$
\begin{equation}\label{QKin two part distr func M}\times\biggl(\delta(\textbf{r}-\textbf{r}_{n})\delta(\textbf{p}-\widehat{\textbf{D}}_{n})+\delta(\textbf{p}-\widehat{\textbf{D}}_{n})\delta(\textbf{r}-\textbf{r}_{n})\biggr)\sigma_{n}^{\alpha}\psi(R,t)+h.c.\Biggr)dR,\end{equation}
and
$$N_{2}^{\alpha}(\textbf{r},\textbf{p},\textbf{r}',\textbf{p}',t)=\frac{1}{4}\int \Biggl(\psi^{*}(R,t)\sum_{n,k\neq n}\delta(\textbf{r}'-\textbf{r}_{k})\delta(\textbf{p}'-\widehat{\textbf{D}}_{k})\times$$
\begin{equation}\label{QKin two part distr func N}\times\biggl(\delta(\textbf{r}-\textbf{r}_{n})\delta(\textbf{p}-\widehat{\textbf{D}}_{n})+\delta(\textbf{p}-\widehat{\textbf{D}}_{n})\delta(\textbf{r}-\textbf{r}_{n})\biggr)\sigma_{k}^{\alpha}\psi(R,t)+h.c.\Biggr)dR.\end{equation}
These two two-particle distribution function describe correlations between spin dynamic of one particle with the space motion (concentration evolution) other particle. Consequently, in the self-consistent field approximation this function represents as
\begin{equation}\label{QKin two part distr func M SCF Appr}M_{2}^{\alpha}(\textbf{r},\textbf{p},\textbf{r}',\textbf{p}',t)=S^{\alpha}(\textbf{r},\textbf{p},t)f(\textbf{r}',\textbf{p}',t),\end{equation}
and
\begin{equation}\label{QKin two part distr func N SCF Appr}N_{2}^{\alpha}(\textbf{r},\textbf{p},\textbf{r}',\textbf{p}',t)=f(\textbf{r},\textbf{p},t)S^{\alpha}(\textbf{r}',\textbf{p}',t).\end{equation}

Using equation (\ref{QKin kinetic equation (spin evol) gen with spin and int with two part F}) we can find hydrodynamical equation for spin evolution $S^{\alpha}(\textbf{r},t)$$=S^{\alpha}(\textbf{r},\textbf{p},t)$. Terms containing $\nabla_{\textbf{p}}f(\textbf{r},\textbf{p},t)$ give no contribution in the hydrodynamical spin evolution equation. Consequently, the First, second, and last terms of equation (\ref{QKin kinetic equation (spin evol) gen with spin and int with two part F}) give contribution in the hydrodynamical spin evolution equation, and we have
$$\partial_{t}S^{\alpha}(\textbf{r},t)+\frac{1}{m}\int \partial_{\textbf{r}}(\textbf{p}S^{\alpha}(\textbf{r},\textbf{p},t))d\textbf{p}$$
\begin{equation}\label{QKin Hydr spin evol}-\frac{2\gamma}{\hbar}\varepsilon^{\alpha\beta\gamma}\Biggl(B^{\beta}S^{\gamma}(\textbf{r},t)+\gamma\int G^{\beta\delta}(\textbf{r},\textbf{r}')S_{2}^{\gamma\delta}(\textbf{r},\textbf{r}',t)d\textbf{r}'\Biggr)=0,\end{equation}
which corresponds to equations presented in Ref.s \cite{Andreev RPJ 07}, \cite{Andreev Arxiv 333}. We admit that we have not introduced velocity field in equation (\ref{QKin Hydr spin evol}). The second term in equation (\ref{QKin Hydr spin evol}) is the divergence of the spin current and the last term describes interactions. Hydrodynamical definition of spin-current we briefly discus  in the Appendix, where we show that presented here definition corresponds to the hydrodynamical one.

The second term in equation (\ref{QKin Hydr spin evol}) is the divergence of the spin-current, and the last term in a big brackets describes interaction. Hydrodynamical spin-current definition we briefly discuss in Appendix.

\subsection{ selfconsistent field approximation for spinning particles}

In previous sections we have considered the self-consistent field approximation for charged spinless particles. After derivation of the general set of kinetic equations for spinning particles we should consider self-consistent field approximation of these equations, to be short, we present equations in the quasi-classic limit. We have presented approximate form for $M_{2}^{\alpha}$ and $N_{2}^{\alpha}$, but we have one more two-particle function $S_{2}^{\alpha\beta}$, which is represented as
\begin{equation}\label{QKin def distribution spin two part function self consist field appr}S_{2}^{\alpha\beta}(\textbf{r},\textbf{p},\textbf{r}',\textbf{p}',t)=S^{\alpha}(\textbf{r},\textbf{p},t)S^{\beta}(\textbf{r}',\textbf{p}',t)\end{equation}
In the result we have next set of equations
\begin{equation}\label{QKin kinetic equation gen  classic limit with E and B}\partial_{t}f+\frac{\textbf{p}}{m}\partial_{\textbf{r}}f+e\biggl(\textbf{E}+\frac{1}{mc}\textbf{p}\times\textbf{B}\biggr)\nabla_{\textbf{p}}f+\partial_{\alpha} B^{\beta}(\textbf{r},t)\partial_{\textbf{p}\alpha} S^{\beta}(\textbf{r}, \textbf{p},t)=0,\end{equation}
and
\begin{equation}\label{QKin kinetic equation gen for spin classic limit with E and B}\partial_{t}S^{\alpha}+\frac{\textbf{p}}{m}\partial_{\textbf{r}}S^{\alpha}+e\biggl(\textbf{E}+\frac{1}{mc}\textbf{p}\times\textbf{B}\biggr)\nabla_{\textbf{p}}S^{\alpha}+\partial_{\gamma} B^{\alpha}(\textbf{r},t)\partial_{\textbf{p}\gamma} f(\textbf{r}, \textbf{p},t)-\frac{2\gamma}{\hbar}\varepsilon^{\alpha\beta\gamma}B^{\beta}S^{\gamma}=0,\end{equation}
Due to the fact that we have considered spin-spin interaction and we have not included spin-current and current-current interaction we have magnetic field satisfying to the following equation $\nabla\times \textbf{B}=4\pi\nabla\times\int \textbf{S}(\textbf{r},\textbf{p},t)d\textbf{p}$. Using additivity of the electromagnetic fields we can include magnetic field caused by the electric currents, therefore we have
\begin{equation}\label{QKin electro stat Max in spin chapter} \begin{array}{ccc}\nabla\times \textbf{E}=0,&   \nabla\textbf{E}=4\pi e\int f(\textbf{r},\textbf{p},t)d\textbf{p}\end{array}\end{equation}
\begin{equation}\label{QKin electro stat Max in spin chapter} \begin{array}{ccc}\nabla\times \textbf{B}=4\pi\nabla\times\int \textbf{S}(\textbf{r},\textbf{p},t)d\textbf{p}+\frac{4\pi}{mc}e\int \textbf{p} f(\textbf{r},\textbf{p},t)d\textbf{p},&   \nabla\textbf{B}=0.\end{array}\end{equation}

Equations (\ref{QKin kinetic equation gen  classic limit with E and B}) and (\ref{QKin kinetic equation gen for spin classic limit with E and B}) are the set of kinetic equations for the quantum plasma of spinning particles. Kinetic equations for quantum spin plasma have been derived in other papers \cite{Brodin PRL 08}, \cite{Marklund TTSP 11}-\cite{Haas NJP 10} However, obtained in this paper equations have several differences with attained before, and main difference is the fact that we get set of two kinetic equation for distribution function $f$ and the spin distribution function $S^{\alpha}$, instead of one equation for the distribution function $f$ only.

\section{Kinetics of particles having electric dipole moment}

The electric dipole moment is the property of neutral and charged particles which is, analogously to the spin, leads to anisotropic inter-particles interaction. Analogously to the spin kinetics we have to have two kinetic equations, one for usual distribution function (\ref{QKin def distribution function}), and another one we can called the dipole distribution function, which is the average of the following operator
\begin{equation}\label{QKin distr dipole func operator} \hat{P}^{\alpha}=\sum_{n}d_{n}^{\alpha}\delta(\textbf{r}-\widehat{\textbf{r}}_{n})\delta(\textbf{p}-\widehat{\textbf{p}}_{n}),\end{equation}
where $d_{n}$ is the electric dipole moment of particle, usually molecules happen to be polar particle, i.e. having electric dipole moment. Quantum hydrodynamics of particles having electric dipole moment was considered in Ref. \cite{Andreev PRB 11}. Boltzmann equation has been used in Ref.s \cite{Chan PRA 10}, \cite{Ronen PRA 10}. As an example of infinite bubble diagram expansion method using for evolution of particles having electric dipole moment see Ref. \cite{Qiuzi Li PRB 11}.
To reveal main features of the electric dipole moment kinetics we present kinetic equations for neutral particles having electric dipole moment.
We start our derivation from the many-particle Schrodinger equation with the Hamiltonian
\begin{equation}\label{QKin dipoles Hamiltonian}\hat{H}=\sum_{i}\Biggl(\frac{1}{2m_{i}}\hat{\textbf{p}}_{i}^{2}-d_{i}^{\alpha}E_{i,ext}^{\alpha}+V_{trap}(\textbf{r}_{i},t)\Biggr)-\frac{1}{2}\sum_{i,j\neq i}\Biggl(d_{i}^{\alpha}d_{j}^{\beta}G_{ij}^{\alpha\beta}\Biggr),\end{equation}
where
the Hamiltonian of the electric dipole interaction  is
\begin{equation}\label{QKin tran Ham dd int}H_{dd}=-\partial^{\alpha}\partial^{\beta}\frac{1}{r}\cdot d_{1}^{\alpha}d_{2}^{\beta},\end{equation}
which corresponds to the Maxwell's equation, as it was shown in Ref. ~\cite{Andreev arxiv 12 02}.

Using well-known identity
\begin{equation}\label{QKin tran togdestvo}-\partial^{\alpha}\partial^{\beta}\frac{1}{r}= \frac{\delta^{\alpha\beta}-3r^{\alpha}r^{\beta}/r^{2}}{r^{3}}+\frac{4\pi}{3}\delta^{\alpha\beta}\delta(\textbf{r}),\end{equation}
we can see that the Hamiltonian (\ref{QKin tran Ham dd int}) differs from usually used one
\begin{equation}\label{QKin tran pot of dd int usual} H_{dd}=\frac{\delta^{\alpha\beta}-3r^{\alpha}r^{\beta}/r^{2}}{r^{3}}d_{1}^{\alpha}d_{2}^{\beta}.\end{equation}
Necessity to consider the Hamiltonian of electric dipole
interaction in the form (\ref{QKin tran Ham dd int}) caused by the
fact that it must accord to the Maxwell's equation. This
connection exists since Maxwell's equations describe electric
field and it's connection with the sources. In our case source is
the density of electric polarization. Action of the electric field
on the density of polarization come in to equations of motion via
the force field. As the result it describes interaction of
polarization (electric dipole moments). To be short we present
them in the self-consistent field approximation
\begin{equation}\label{QKin kinetic equation gen classic limit dip chapter with E and B}
\partial_{t}f+\frac{\textbf{p}}{m}\partial_{\textbf{r}}f+e\biggl(\textbf{E}+\frac{1}{mc}\textbf{p}\times\textbf{B}\biggr)\nabla_{\textbf{p}}f+\partial_{\alpha} B^{\beta}(\textbf{r},t)\partial_{\textbf{p}\alpha} P^{\beta}(\textbf{r}, \textbf{p},t)=0,\end{equation}
and
\begin{equation}\label{QKin kinetic equation gen for dip classic limit  dip chapter with E and B}
\partial_{t}P^{\alpha}+\frac{\textbf{p}}{m}\partial_{\textbf{r}}P^{\alpha}+e\biggl(\textbf{E}+\frac{1}{mc}\textbf{p}\times\textbf{B}\biggr)\nabla_{\textbf{p}}P^{\alpha}+\partial_{\gamma} E^{\beta}(\textbf{r},t)\partial_{\textbf{p}\gamma} R^{\alpha\beta}(\textbf{r}, \textbf{p},t)=0,\end{equation}
where
$$R^{\alpha\beta}(\textbf{r}, \textbf{p},t)=\frac{1}{4}\int\Biggl(\psi^{*}\sum_{n}d_{n}^{\alpha}d_{n}^{\beta}(\delta(\textbf{r}-\widehat{\textbf{r}}_{n})\delta(\textbf{p}-\widehat{\textbf{p}}_{n})+\delta(\textbf{p}-\widehat{\textbf{p}}_{n})\delta(\textbf{r}-\widehat{\textbf{r}}_{n}))+c.c.\Biggr)dR$$
is the one particle distribution function depending of product of the two operators of the electric dipole moment of one particle, and the electric field caused inter-particle interaction satisfy
to the Maxwell equations
\begin{equation}\label{QKin electro stat Max el dip} \begin{array}{ccc}\nabla\times \textbf{E}=0,&   \nabla\textbf{E}=-4\pi \nabla\int \textbf{P}(\textbf{r},\textbf{p},t)d\textbf{p}.\end{array}\end{equation}

At development of the quantum hydrodynamical description of
systems of particles having electric dipole moment \cite{Andreev
PRB 11} has been suggested a method of closing of the QHD set.
Analogous approximation can be used here.

\section{Kinetics of neutral particles with short range interaction}

Presented derivation can be used for systems of particles with the
short-range interaction. The quantum hydrodynamics of neutral
particles with the short-range interaction was considered in Ref.
\cite{Andreev PRA08}, where ultracold quantum gases were
considered particularly. Following ideas developed in this paper
and in Ref. \cite{Andreev PRA08} we can derive kinetic equation
for described systems. To be short we present corresponding
kinetic equation in the quasi-classical limit
\begin{equation}\label{QKin kinetic equation gen SRI classic limit}\partial_{t}f+\frac{\textbf{p}}{m}\partial_{\textbf{r}}f-\nabla V_{ext}\nabla_{\textbf{p}}f-
\int \nabla_{\textbf{r}}U(\textbf{r}-\textbf{r}')\cdot\nabla_{\textbf{p}} f_{2}(\textbf{r},\textbf{p},\textbf{r}',\textbf{p}',t) d\textbf{r}'d\textbf{p}'
=0,\end{equation}
using the fact that we consider the short-range interaction we can rewrite general equation (\ref{QKin kinetic equation gen SRI classic limit}) in the following form
\begin{equation}\label{QKin kinetic equation gen SRI FOIR classic limit}\partial_{t}f+\frac{\textbf{p}}{m}\partial_{\textbf{r}}f-\nabla V_{ext}\nabla_{\textbf{p}}f+\frac{1}{2}\Upsilon\partial_{\textbf{r}}\partial_{\textbf{p}}\int f_{2}(\textbf{r}, \textbf{p}, \textbf{r}, \textbf{p}',t) d\textbf{p}'=0,\end{equation}
where
\begin{equation}\label{Upsilon} \Upsilon=\frac{4\pi}{3}\int
dr(r)^{3}\frac{\partial U(r)}{\partial r},\end{equation}
and we have made additional approximation, we have considered the short-range interaction in the first order by the interaction radius \cite{Andreev PRA08}. In general case the short-range interaction leads to the expansion of the term describing interaction in the series, and we have included only first term of this expansion. This scheme was developed in quantum hydrodynamics \cite{Andreev PRA08}, contribution of the next nonzero term in this expansion, in hydrodynamical approximation, can be found in Ref.s \cite{Andreev PRA08}, \cite{Andreev Izv.Vuzov. 09 1}, \cite{Andreev MPL 12}, \cite{Zezyulin arxiv 12}.

Integrating the last term in equation (\ref{QKin kinetic equation gen SRI FOIR classic limit}) gives us trace of hydrodynamical two-particle concentration at great length considered in Ref.s \cite{Andreev PRA08}, \cite{Andreev Izv.Vuzov. 09 1}
$$n_{2}(\textbf{r},\textbf{r},t)\equiv Tr n_{2}(\textbf{r},\textbf{r}',t)=\int f_{2}(\textbf{r}, \textbf{p}, \textbf{r}, \textbf{p}',t) d\textbf{p}d\textbf{p}'.$$

Trace of the two-particle concentration was calculated in Ref.
\cite{Andreev PRA08}, where was shown, in the first order by the
interaction radius, that
$$n_{2}(\textbf{r},\textbf{r},t)=n^{2}(\textbf{r},t),$$
for system of particles being in the Bose-Einstein condensate
state.

Methods developed in Ref. \cite{Andreev PRA08} can be used for
further studying of the structure of the quantum kinetic equation
for quantum gases.

\section{Graphene kinetics}

This section is dedicated to development of the quantum kinetic
theory for graphene carriers. It is important to mention several
interesting papers have been recently published, where possible
applications of graphene for future devises is discussed
\cite{Pasanen PS 12}, \cite{Bae PS 12} and history of graphene
studies is presented \cite{Geim PS 12}.

Methods of physical kinetics are usually used for studying of
collective properties of many-particle systems, such as dispersion
dependencies of collective excitations \emph{and} their linear and
nonlinear evolution. Properties of collective excitations in
graphene, including spectrum of linear plasmons, were considered
in Ref.s \cite{Sarma PRL 09}, \cite{Jablan PRB 09},
\cite{Falkovsky EPJ B 07}, \cite{Dubinov JP C 11} by means of the
random phase approximation (RPA).

Working with the graphene we have deal with the system of charged
spinning particles in the presence, in general case, of an
external magnetic field. Therefore we have to use the distribution
function defined by formula (\ref{QKin def distribution function
in magn field}).  Before derivation of the kinetic equation for
graphene we will describe basic equation which drives graphene
electrons in the vicinity of the Dirac points.

We use the
many-particle spinor massless Dirac equation ~\cite{Sheehy PRL 07},
~\cite{Novoselov nature 05}
\begin{equation}\label{QKin GR Hamiltonian}\imath\hbar\partial_{t}\psi=\Biggl(\sum_{i}\biggl(v_{F}\sigma^{\alpha}\hat{D}_{i}^{\alpha}+e_{i}\varphi_{i,ext}\biggr)+\frac{1}{2}\sum_{i,j\neq i}e_{i}e_{j}G_{ij}\Biggr)\psi.\end{equation}
The following designations are used in the Hamiltonian (\ref{QKin GR Hamiltonian}):
$D_{i}^{\alpha}=-\imath\hbar\partial_{i}^{\alpha}-e_{i}A_{i,ext}^{\alpha}/c$,
 $\varphi_{i,ext}$, $A_{i,ext}^{\alpha}$ - is the potentials
of the external electromagnetic field,
$\textbf{E}_{i,ext}=-\nabla\varphi_{i,ext}-\partial_{t}\textbf{A}_{i,ext}$
is the electric field, $\textbf{B}_{i,ext}=curl\textbf{A}_{i,ext}$
is the magnetic field, quantities $e_{i}$, $m_{i}$-are the charge
and mass of particles, $\hbar$-is the Planck constant, and
$G_{ij}=1/r_{ij}$ is the Green functions of the Coulomb
interaction. In equation
(\ref{QKin GR Hamiltonian}) the spinor wave function
$\psi=\psi(R,t)$ depend on 2N coordinates $R=[\textbf{r}_{1}, ...,
\textbf{r}_{N}]$ and time, where $\textbf{r}_{i}=[x_{i}, y_{i}]$
is the 2D coordinates of each particle. Potentials
$\varphi_{i,ext}=\varphi_{ext}(\textbf{r}_{i},t)$,
$A_{i,ext}^{\alpha}=A_{ext}^{\alpha}(\textbf{r}_{i},t)$ also
depend on 2D variables. This fact has deep consequences.
Potential part of electric field connected with the scalar
potential via space derivative:
$\textbf{E}_{i}=-\nabla_{i}\varphi_{i}$. Consequently in equation
(\ref{QKin GR Hamiltonian}) there is no contribution of external
electric field directed perpendicular to the graphene plane $E_{z}$ (Contribution of $E_{z}$ might
appear via $\partial_{t}A_{z}$). Physically, there is no
limitation on attendance of $z$ projection of electric field and
it's action on graphene electrons. The magnetic field vector to be
$$\textbf{B}=curl\textbf{A}=\textbf{e}_{x}(\partial_{y}A_{z}-\partial_{z}A_{y})$$
$$+\textbf{e}_{y}(\partial_{z}A_{x}-\partial_{x}A_{z})+\textbf{e}_{z}(\partial_{x}A_{y}-\partial_{y}A_{x}),$$
two component of the vector potential of the magnetic field
$A_{x}$, $A_{y}$are presented in Hamiltonian (\ref{QKin GR
Hamiltonian}), and  they does not depend on coordinate $z$.
Therefore equation (\ref{QKin GR Hamiltonian}) contain $z$ component
of the magnetic field only. In this paper we interested in action of the
external magnetic field directed at angle of graphene plane.
Therefore, we get graphene kinetic equations including whole vector of
magnetic field $B\cdot e_{z}\rightarrow
\textbf{B}=[B_{x},B_{y},B_{z}]$.

Quantum hydrodynamics corresponding to described microscopic
theory was developed in Ref. \cite{Andreev arxiv 12 01}, \cite{Andreev PIERS 2012 graphene}.
Kinetic equation for graphene was used in Ref. \cite{Svintsov arxiv 12}
for derivation of the hydrodynamic equations.

It should be admitted that the spin matrix vector
$\sigma^{\alpha}$ describe not spin, but so called quasi-spin.
This notion appears in graphene due to the fact that that Wigner-Seitz cell
lattice contains two carbon atoms. Thus, graphene lattice consists
of two sub-lattices, which are usually called lattice  A and
lattice B. Therefore, in his motion the graphene $p_{Z}$ electron
moves from atom A (B) to atom B (A), and transitions of the
graphene electron between sub-lattices A and B is modelled by
means quasi-spin.

Kinematic term in equation (\ref{QKin GR Hamiltonian}) is the
helicity of particles $\sigma^{\alpha} p^{\alpha}$, so we have
here that the Hamiltonian is proportional to the helicity of
particles, and we consider dynamics governed by the spin 1/2 two
dimensional helicity as the kinetic energy of particle. In Ref.
\cite{Andreev arxiv 12 GrExcBEC} we have considered the QHD
formulation of the particle dynamics governed by the spin-1 two
dimensional helicity. It has been suggested that such equations
might describe evolution of excitons in the graphene, existing in
two layer graphene. In the next section  we will consider kinetic
formulation of this model. Here we continue development of the
kinetic theory for electron in monolayer graphene.

Differentiating distribution function (\ref{QKin def distribution function in magn field}) with respect to time and using equation (\ref{QKin GR Hamiltonian}) for time derivatives of the wave function we find the graphene kinetic equation. In general case it has form
$$\partial_{t}f+v_{F}\nabla\textbf{S}-\frac{e}{c}\partial_{t}\textbf{A}\nabla_{\textbf{p}}f-v_{F}\frac{e}{c}\partial^{\alpha}A^{\beta}\partial^{\beta}_{\textbf{p}}S^{\alpha}$$
$$-v_{F}\frac{e}{c}\frac{\imath}{\hbar}A^{\alpha}\sin(\overleftarrow{\nabla}_{\textbf{r}}\nabla_{p})S^{\alpha}+e\frac{\imath}{\hbar}\varphi\sin(\overleftarrow{\nabla}_{\textbf{r}}\nabla_{\textbf{p}})f$$
\begin{equation}\label{QKin GR kin eq gen}+e^{2}\frac{\imath}{\hbar}\int G(\textbf{r},\textbf{r}')\sin(\overleftarrow{\nabla}_{\textbf{r}}\nabla_{\textbf{p}})f_{2}(\textbf{r},\textbf{p}, \textbf{r}', \textbf{p}', t)d\textbf{r}'d\textbf{p}'=0.\end{equation}
This equation contains the spin distribution function
$S^{\alpha}(\textbf{r},\textbf{p},t)$, so we will derive equation
for the spin distribution function
$S^{\alpha}(\textbf{r},\textbf{p},t)$. As well, we rewrite
equation (\ref{QKin GR kin eq spin gen}) in the classical-like
limit
$$\partial_{t}f+v_{F}\nabla\textbf{S}+e\textbf{E}\nabla_{\textbf{p}}f-v_{F}\frac{e}{c}\varepsilon^{\alpha\beta\gamma}B^{\gamma}\partial_{\textbf{p}}^{\beta}S^{\alpha}$$
\begin{equation}\label{QKin GR kin eq class like}-e^{2}\int\partial^{\alpha}G(\textbf{r},\textbf{r}')\partial^{\alpha}_{\textbf{p}}f_{2}(\textbf{r},\textbf{p}, \textbf{r}', \textbf{p}', t)d\textbf{r}'d\textbf{p}'=0.\end{equation}
As it discussed in Ref. \cite{Sarma PRL 09}, graphene properties
have no classic limit, and the notion "classical-like limit" is
used here to show that we consider the first term of operator
$\sin(\overleftarrow{\nabla}_{\textbf{r}}\nabla_{p})$ only.
Obtained equation (\ref{QKin GR kin eq class like}) is quantum
equation, but it accounts leading on $\hbar$ terms only.

We have to admit that vectors $\textbf{r}$ and $\textbf{p}$ are
two-dimensional vectors, however vectors
$\textbf{E}(\textbf{r},t)$ and $\textbf{B}(\textbf{r},t)$, and
even pseudo-spin vector $\textbf{S}(\textbf{r},t)$ can be
considered as three dimensional vectors. Belonging of the graphene
electrons to the lattice A and B is modelled by increasing of
amplitude of $\psi_{A}$ or $\psi_{B}$, where $\psi^{*}=(\psi_{A},
\psi_{B})$ spinor function of electron, and appearing of the
$S_{z}$ during derivative at using of the commutation relation for
spin matrices is quite reasonable. Since belonging of the electron
to lattice A or B describes by the structure of the Pauli
matrices, and we can easily use some combination of the Pauli
matrices as they appear at derivation.

Kinetic equation for the spin distribution function appears as
$$\partial_{t}S^{\alpha}+v_{F}\partial_{\textbf{r}}^{\alpha}f+\frac{2v_{F}}{\hbar}\varepsilon^{\alpha\beta\gamma}p^{\gamma}S^{\beta}-\frac{e}{c}\partial_{t}A^{\beta}\partial_{\textbf{p}}^{\beta}S^{\alpha}$$
$$-v_{F}\frac{e}{c}\partial^{\alpha}A^{\beta}\partial^{\beta}_{\textbf{p}}f+v_{F}\frac{e}{c}\frac{\imath}{\hbar}A^{\alpha}\sin(\overleftarrow{\nabla}_{\textbf{r}}\nabla_{\textbf{p}})f+e\frac{\imath}{\hbar}\varphi\sin(\overleftarrow{\nabla}_{\textbf{r}}\nabla_{\textbf{p}})S^{\alpha}$$
\begin{equation}\label{QKin GR kin eq spin gen}+e^{2}\frac{\imath}{\hbar}\int G(\textbf{r},\textbf{r}')\sin(\overleftarrow{\nabla}_{\textbf{r}}\nabla_{\textbf{p}})M_{2}^{\alpha}(\textbf{r},\textbf{p},\textbf{r}',\textbf{p}',t)d\textbf{r}'d\textbf{p}'=0,\end{equation}
and in the classical like limit it has form
$$\partial_{t}S^{\alpha}+v_{F}\partial_{\textbf{r}}^{\alpha}f+\frac{2v_{F}}{\hbar}\varepsilon^{\alpha\beta\gamma}p^{\gamma}S^{\beta}+eE^{\beta}\partial_{\textbf{p}}^{\beta}S^{\alpha}
-v_{F}\frac{e}{c}\varepsilon^{\alpha\beta\gamma}B^{\gamma}\partial^{\beta}_{\textbf{p}}f$$
\begin{equation}\label{QKin GR kin spin eq class like}-e^{2}\int \nabla_{\textbf{r}}G(\textbf{r},\textbf{r}')\nabla_{\textbf{p}}M_{2}^{\alpha}(\textbf{r},\textbf{p},\textbf{r}',\textbf{p}',t)d\textbf{r}'d\textbf{p}'=0.\end{equation}

In the absence of inter-particle interaction we can rewrite the couple of graphene kinetic equations
\begin{equation}\label{QKin GR kin eq class like without int}\partial_{t}f+v_{F}\nabla\textbf{S}+e\textbf{E}\nabla_{\textbf{p}}f-v_{F}\frac{e}{c}\varepsilon^{\alpha\beta\gamma}B^{\gamma}\partial_{\textbf{p}}^{\beta}S^{\alpha}=0,\end{equation}
and
\begin{equation}\label{QKin GR kin spin eq class like without int}\partial_{t}S^{\alpha}+v_{F}\partial_{\textbf{r}}^{\alpha}f+\frac{2v_{F}}{\hbar}\varepsilon^{\alpha\beta\gamma}p^{\gamma}S^{\beta}+e\textbf{E} \nabla_{\textbf{p}}S^{\alpha}-v_{F}\frac{e}{c}\varepsilon^{\alpha\beta\gamma}B^{\gamma}\partial^{\beta}_{\textbf{p}}f=0.\end{equation}

In the self-consistent field approximation the graphene kinetic
equation (\ref{QKin GR kin eq class like}), (\ref{QKin GR kin spin
eq class like}) transforms in
$$\partial_{t}f+v_{F}\nabla\textbf{S}+e\textbf{E}\nabla_{\textbf{p}}f-v_{F}\frac{e}{c}\varepsilon^{\alpha\beta\gamma}B^{\gamma}\partial_{\textbf{p}}^{\beta}S^{\alpha}$$
\begin{equation}\label{QKin GR kin eq class like self-consist}-e^{2}\int\partial^{\alpha}G(\textbf{r},\textbf{r}')f(\textbf{r}', \textbf{p}', t)d\textbf{r}'d\textbf{p}' \cdot\partial^{\alpha}_{\textbf{p}}f(\textbf{r},\textbf{p}, t)=0,\end{equation}
and
$$\partial_{t}S^{\alpha}+v_{F}\partial_{\textbf{r}}^{\alpha}f+\frac{2v_{F}}{\hbar}\varepsilon^{\alpha\beta\gamma}p^{\gamma}S^{\beta}+e\textbf{E} \nabla_{\textbf{p}}S^{\alpha}-v_{F}\frac{e}{c}\varepsilon^{\alpha\beta\gamma}B^{\gamma}\partial^{\beta}_{\textbf{p}}f$$
\begin{equation}\label{QKin GR kin spin eq class like self-consist}-e^{2}\int \nabla_{\textbf{r}}G(\textbf{r},\textbf{r}')f(\textbf{r}',\textbf{p}',t)d\textbf{r}'d\textbf{p}'  \cdot\nabla_{\textbf{p}}S^{\alpha}(\textbf{r},\textbf{p},t)=0.\end{equation}
Due to two-dimensionality of the system we can not introduce the
field of interaction, or if we actually would be forced to do it
we get $\delta$ function in the right-hand side of the field
equation $\nabla \textbf{E}$$=4\pi\rho$$=4\pi e \delta(z)\int
f(\textbf{r},\textbf{p},t)d^{2}\textbf{p}$.

In the result we have the closed set of quantum kinetic equations
for graphene, where evolution of the two distribution functions
obeys to the two kinetic equations, see for example (\ref{QKin GR
kin eq class like self-consist} and (\ref{QKin GR kin spin eq
class like self-consist})).

\section{Graphene exciton kinetics}

In the previous section we have considered kinetic equation for
carriers in graphene, which can be considered as a spin-1/2
helicity governed system of charged particles. In this section we
going to study kinetic properties of the spin-1 helicity governed
particles motion. This model has been suggested in Ref.
\cite{Andreev arxiv 12 GrExcBEC} for obtaining of the quantum
hydrodynamical descriptions of the graphene excitons  and
dispersion dependence of collective excitations of the
Bose-Einstein condensate of the graphene excitons. Here we present
corresponding kinetic equations.

We present a basic equation here for evolution description of
excitons in graphene
\begin{equation}\label{QKin Hamiltonian helicity S1}\imath\hbar\partial_{t}\psi=\Biggl(\sum_{i}\biggl(v_{F}\hat{s}^{\alpha}_{i}p_{i}^{\alpha}+V_{i,ext}\biggr)+\frac{1}{2}\sum_{i,j\neq i}U_{ij}\Biggr)\psi.\end{equation}
Using this equation we derive and present below the equations for
description of excitons collective motion. Equation (\ref{QKin
Hamiltonian helicity S1}) differs from (\ref{QKin GR Hamiltonian})
by the form of spin matrix and the form of interaction. Equation
(\ref{QKin Hamiltonian helicity S1}) contains following
quantities: wave function $\psi=\psi(R,t)$, $R$ is the whole
particles coordinates $R=[\textbf{r}_{1}, ..., \textbf{r}_{i},
..., \textbf{r}_{N}]$, $\textbf{r}_{i}=[x_{i},y_{i}]$,
$\hat{s}^{\alpha}_{i}$ are the spin-1 matrixes for $i$-th
particle, $p_{i}^{\alpha}=-\imath\hbar\nabla$ is the momentum
operator, $V_{i,ext}$ is the potential of external field, $U_{ij}$
is the short-range interaction potential describing the
interaction between excitons in graphene. We notice that at the
same time some particle might interact with several particles, by
means short-range interaction potential $U_{ij}$. For this
statement illustration we refer to the liquid where molecules is
neutral and interacts with the several neighbor molecules. We
consider quasi spin-1 particles and spin operators are $3\times3$
matrixes
$$\begin{array}{ccc} \hat{s}_{x}=\frac{1}{\sqrt{2}}\left(\begin{array}{ccc}0&
1&
0\\
1&
0&
1\\
0&
1&
0\\
\end{array}\right),&
\hat{s}_{y}=\frac{1}{\sqrt{2}}\left(\begin{array}{ccc}0& -\imath &
0\\
\imath &
0&
-\imath \\
0&
\imath &
0\\
\end{array}\right),&
\hat{s}_{z}=\left(\begin{array}{ccc}1& 0&
0\\
0& 0&
0\\
0& 0&
-1\\
\end{array}\right),\end{array}$$
and the commutation relation for spin-1 matrixes is
\begin{equation}\label{QKin comm rel spin 1} [\hat{s}^{\alpha}_{i},\hat{s}^{\beta}_{j}]=\imath\delta_{ij}\varepsilon^{\alpha\beta\gamma}\hat{s}^{\gamma}_{i}.\end{equation}

Differentiating the distribution function , defined by formula
(\ref{QKin def distribution function in magn field}), but without
vector potential, with respect to time we find following kinetic
equation
\begin{equation}\label{QKin GR excit}\partial_{t}f+v_{F}\nabla\textbf{S}-\nabla V_{ext}\nabla_{\textbf{p}}f+\frac{1}{2}\Upsilon_{2D}\partial^{\alpha}\partial^{\alpha}_{\textbf{p}}\int f_{2}(\textbf{r},\textbf{p}, \textbf{r}, \textbf{p}', t)d\textbf{p}'=0.\end{equation}
which contains quasi-spin distribution function (\ref{QKin def
spin distribution function}), where we have operators
$\hat{s}^{\alpha}_{n}$ instead of the Pauli matrices
$\widehat{\sigma}^{\alpha}_{n}$, and, in the same way, we obtain
kinetic equation for the spin-distribution function
$$\partial_{t}S^{\alpha}+v_{F}\partial_{\textbf{r}}^{\beta}C^{\alpha\beta}+\frac{v_{F}}{\hbar}\varepsilon^{\alpha\beta\gamma}p^{\gamma}S^{\beta}-\nabla^{\beta} V_{ext}\partial_{\textbf{p}}^{\beta}S^{\alpha}$$
\begin{equation}\label{QKin GR excit for S}+\frac{1}{2}\Upsilon_{2D}\nabla_{\textbf{r}}\nabla_{\textbf{p}}\int M_{2}^{\alpha}(\textbf{r},\textbf{p},\textbf{r},\textbf{p}',t)d\textbf{p}'=0.\end{equation}
where
$$C^{\alpha\beta}(\textbf{r}, \textbf{p},t)=\frac{1}{4}\int
\Biggl(\psi^{*}(R,t)\sum_{n}\biggl(\delta(\textbf{r}-\textbf{r}_{n})\delta(\textbf{p}-\widehat{\textbf{D}}_{n})+\delta(\textbf{p}-\widehat{\textbf{D}}_{n})\delta(\textbf{r}-\textbf{r}_{n})\biggr)\times$$
\begin{equation}\label{QKin def spin C distribution function}\times\sigma^{\alpha}_{n}\sigma^{\beta}_{n}\psi(R,t)+h.c.\Biggr)dR,\end{equation}
and
\begin{equation}\label{QKin Upsilon 2D} \Upsilon_{2D}=\pi\int dr(r)^{2}\frac{\partial U(r)}{\partial r}. \end{equation}
$\Upsilon_{2D}$ is the interaction constant for short-range
interaction in two-dimensional system. We have obtained additional
one particle function $C^{\alpha\beta}$, which unlike spin-1/2
case does not represents via $f$ and $S^{\alpha}$. As for
electrically polarized particles, in section (V), appearing of
additional function make truncation more difficult. We have
suggested some approximation for hydrodynamical description
\cite{Andreev PRB 11}, \cite{Andreev arxiv 12 02}, \cite{Andreev
arxiv 12 GrExcBEC}, where we have considered collective wave
dispersion. We can suggest analogous approximations, but
developing method allows us to perform truncation using
approximate form of the wave function.

\section{Conclusion}

Methods of studying of the quantum plasma have been developing last decade. The task of the interparticle interaction account reveals as a nontrivial
problem. However we can conclude that this problem has been successfully solve in nonrelativistic in semi-relativistic quantum hydrodynamics. In classical physics kinetics method is more popular and profound method of the plasma studying. Thus it is worthwhile to derive the quantum kinetic theory, and it is also important to trace the interparticle interaction from microscopic level, as it was made in quantum hydrodynamics. First of all the kinetics allows to consider thermal effects. In most cases, when we have deal with the quantum plasma, quantum properties might reveal at low temperatures. Thus using of the hydrodynamics in quantum case is more useful than in classical one. Nevertheless a some thermal spreading might leads to new effects in quantum plasma, and we have to have quantum kinetic theory to get correct description.

This program has been realized in the paper. We have obtained quantum kinetic description of different physical systems and presented corresponding kinetic equations in various approximations.

We have suggested new definition for the quantum distribution function which gives us possibility to derive a chain of the kinetics equations directly from many-particle Schrodinger equation including explicitly inter-particle interaction. This derivation is the direct generalization of the many-particle quantum hydrodynamic method. This definition changes at account of the vector potential of the external electromagnetic field by evident explicit replacement of the momentum operator $\widehat{\textbf{p}}_{n}=-\imath\hbar\nabla_{n}$ on $\widehat{\textbf{D}}_{n}=-\imath\hbar\nabla_{n}-e_{n}\textbf{A}_{n}/c$. In the quasi-classical limit it gives the Lorentz force in the usual form, and it also corresponds to the quantum  hydrodynamics of spinning particles.
Developed description corresponds to the Vlasov equation in the classical limit.

We have derived kinetic equations for charged spinning particles (quantum plasma), neutral particles with short-range interaction, electrons in graphene, electrically polarized molecules, and graphene excitons. We have shown that in the system of spinning particles  we have to use set of two kinetic equations, one for usual distribution function and the another for spin-distribution function which arise in the equation of distribution function evolution. We have obtained closed set of kinetic equation which allow to use them for studying of particular problems.

\appendix
\section{}

Here we will proof following formula (\ref{QKin current}), which we rewrite here
\begin{equation}\label{QKin current ap}\textbf{j}(\textbf{r},t)=\int \textbf{p}f(\textbf{r}, \textbf{p},t) d\textbf{p},\end{equation}
we will make including presence of an external magnetic field, that corresponds to the fact that we should use general definition of the distribution function (\ref{QKin def distribution function in magn field}). Distribution function $f$ defined by formula (\ref{QKin def distribution function in magn field}), or in the absence of the magnetic field and the rotational electric field the distribution function has more simple form (\ref{QKin def distribution function}). Hydrodynamical current $\textbf{j}(\textbf{r},t)$ has form \cite{MaksimovTMP 1999}, \cite{MaksimovTMP 2001}, \cite{Andreev RPJ 07}, \cite{Andreev PRB 11}, \cite{Andreev Arxiv 333}
\begin{equation}\label{QKin hydr current ap}\textbf{j}(\textbf{r},t)=\int \sum_{n=1}^{N}\frac{1}{2}\delta(\textbf{r}-\textbf{r}_{n})\biggl( (\widehat{\textbf{p}}\psi)^{*}\psi+\psi^{*}\widehat{\textbf{p}}\psi\biggr) dR.\end{equation}
Now we are ready to demonstration of the formula (\ref{QKin current}) or (\ref{QKin current ap}).
$$\int \textbf{p}f(\textbf{r}, \textbf{p},t) d\textbf{p}=$$
$$=\frac{1}{4}\int \textbf{p} \int \sum_{n=1}^{N}\Biggl(\psi^{*}\biggl(\delta(\textbf{r}-\textbf{r}_{n})\delta(\textbf{p}-\widehat{\textbf{p}}_{n})+\delta(\textbf{p}-\widehat{\textbf{p}}_{n})\delta(\textbf{r}-\textbf{r}_{n})\biggr)\psi +c.c.\Biggr) d\textbf{p}$$
$$=\frac{1}{4}\int \textbf{p} \int \sum_{n=1}^{N}\Biggl(\psi^{*}\biggl(\delta(\textbf{r}-\textbf{r}_{n})\delta(\textbf{p}-\widehat{\textbf{p}}_{n})+\delta(\textbf{p}-\widehat{\textbf{p}}_{n})\delta(\textbf{r}-\textbf{r}_{n})\biggr)\psi$$
$$+\biggl(\delta(\textbf{r}-\textbf{r}_{n})\delta(\textbf{p}-\widehat{\textbf{p}}_{n})+\delta(\textbf{p}-\widehat{\textbf{p}}_{n})\delta(\textbf{r}-\textbf{r}_{n})\biggr)\psi^{*}\cdot\psi\Biggr) d\textbf{p}$$
$$=\frac{1}{4}\int  \int \sum_{n=1}^{N}\Biggl(\psi^{*}\biggl(\delta(\textbf{r}-\textbf{r}_{n})\delta(\textbf{p}-\widehat{\textbf{p}}_{n})+\delta(\textbf{p}-\widehat{\textbf{p}}_{n})\delta(\textbf{r}-\textbf{r}_{n})\biggr)\textbf{p}\psi$$
$$+\biggl(\delta(\textbf{r}-\textbf{r}_{n})\delta(\textbf{p}-\widehat{\textbf{p}}_{n})+\delta(\textbf{p}-\widehat{\textbf{p}}_{n})\delta(\textbf{r}-\textbf{r}_{n})\biggr)(\textbf{p}\psi)^{*}\cdot\psi\Biggr) d\textbf{p}$$
Integrating on momentum $\textbf{p}$ we replace $\textbf{p}$ on $\widehat{\textbf{p}}_{n}$ due to the delta functions $\delta(\textbf{p}-\widehat{\textbf{p}}_{n})$ and find
$$\int \textbf{p}f(\textbf{r}, \textbf{p},t) d\textbf{p}=\frac{1}{2}\int   \sum_{n=1}^{N}\frac{1}{2}\Biggl((\widehat{\textbf{p}}_{n}\psi)^{*}\biggl(\delta(\textbf{r}-\textbf{r}_{n})+\delta(\textbf{r}-\textbf{r}_{n})\biggr)\psi$$
$$+\psi^{*}\biggl(\delta(\textbf{r}-\textbf{r}_{n})+\delta(\textbf{r}-\textbf{r}_{n})\biggr)\widehat{\textbf{p}}_{n}\psi\Biggr) d\textbf{p}=\textbf{j}(\textbf{r},t)$$

Analogously, in the presence of the external magnetic field, we
can show that
\begin{equation}\label{QKin current ap}\textbf{j}(\textbf{r},t)=\int \textbf{p}f(\textbf{r}, \textbf{p},t) d\textbf{p},\end{equation}
where, from hydrodynamic point of view
\begin{equation}\label{QKin hydr current ap}\textbf{j}(\textbf{r},t)=\int \sum_{n=1}^{N}\frac{1}{2}\delta(\textbf{r}-\textbf{r}_{n})\biggl( (\widehat{\textbf{D}}\psi)^{*}\psi+\psi^{*}\widehat{\textbf{D}}\psi\biggr) dR.\end{equation}

In formula (\ref{QKin kin energy}) the hydrodynamic kinetic energy was mentioned, it's explicit form is
\begin{equation}\label{QKin hydr kin energy ap}\varepsilon(\textbf{r},t)=\int \sum_{n=1}^{N}\frac{1}{4m_{n}}\delta(\textbf{r}-\textbf{r}_{n})\biggl( (\widehat{\textbf{D}}^{2}\psi)^{*}\psi+\psi^{*}\widehat{\textbf{D}}^{2}\psi\biggr) dR.\end{equation}
Formula (\ref{QKin kin energy}) can be proved analogously to the proving of formula (\ref{QKin current}) or (\ref{QKin current ap}).

The hydrodynamical spin-current appears in the formula (\ref{QKin Hydr spin evol}), where the second term we can rewrite as
\begin{equation}\label{QKin } \frac{1}{m}\int \partial_{\textbf{r}}^{\beta}(p^{\beta}S^{\alpha}(\textbf{r},\textbf{p},t))d\textbf{p}=\partial^{\beta}\frac{1}{m}\int p^{\beta}S^{\alpha}(\textbf{r},\textbf{p},t)d\textbf{p}\equiv\partial^{\beta}J^{\alpha\beta}(\textbf{r},t),\end{equation}
where
\begin{equation}\label{QKin hydr spin current definition}J^{\alpha\beta}(\textbf{r},t)=\int \sum_{n}\frac{1}{2}\delta(\textbf{r}-\textbf{r}_{n})\biggl( (\widehat{D}^{\beta}\psi)^{*}\sigma^{\alpha}\psi+\psi^{*}\widehat{D}^{\beta}\sigma^{\alpha}\psi\biggr) dR .\end{equation}
It can be proved that formula (\ref{QKin hydr spin current definition}) corresponds to the QHD definition of the spin-current \cite{MaksimovTMP 2001}, \cite{Andreev RPJ 07}, \cite{Andreev arxiv 12 SpinCurrent}.

\subsection{discussion of some methods derivation of kinetic equations}

In nineteen century, Boltzmann suggested his famous kinetic
equation to describe an evolution of a gas of interacting
particles. To describe interaction he introduced the collision
integral. Since then the idea of using collision integral become
widely used. It has been used even for plasma description, along
with self-consistent field. Fundamental meaning of the collision
integral we can understand by reading famous Landau's course of
theoretical physics \cite{Landau Vol 9}. In last decade a
Boltzmann type equation has been used to description of ultracold
quantum gases to describe a part "existing in excited states" that
corresponds to the part of particles at non-zero temperature
\cite{Nikuni PRA 01R}-\cite{Nikuni PRA 01}. A derivation of
kinetic equation from NLSE was made in Ref. \cite{Marklund EPJ B
05} using Wigner distribution function.

Let's pay attention to the ideas presented in Landau's course of
theoretical physics \cite{Landau Vol 9}. At description of
non-equilibrium states of a Fermi- or Bose-liquid one uses
quasi-particle distribution function $\hat{n}$. This function
depends on momentum, coordinate and time, and satisfies a
transport (kinetic) equation. One takes the transport equation in
the form of
\begin{equation}\label{QKin kinet eq with coll int}\frac{d\hat{n}(\textbf{r},\textbf{p},t) }{dt} =I(\hat{n}),\end{equation}
where $I(n)$ is the collision integral and $d/dt$ is an operator which has following form:
$$\frac{d}{dt}=\frac{\partial}{\partial t}+\widehat{\dot{\textbf{r}}} \frac{\partial }{\partial \textbf{r}}+ \widehat{\dot{\textbf{p}}}\frac{\partial}{\partial\textbf{p}},$$
where $\dot{\textbf{r}}$ and $\dot{\textbf{p}}$ should be find from Hamilton's equations.
%this idea have been used in \cite{Brodin PRL 08} for spinning particles

Using of such representation to find an evident form of a
transport equation is not correct, because transport equation is
specified in six dimensional space, where $\textbf{r}$,
$\textbf{p}$ and $t$ are independent variables \emph{and}
quantities $\dot{\textbf{r}}$ and $\dot{\textbf{p}}$ are not
specified and have no sense.

Consideration of transport equation as a continuity equation in
six dimensional phase space was very useful, it was a way of
understanding of meaning of transport equation by many physicists.
But kinetics chains have structure different from structure of
hydrodynamics one.

We have made criticism of the way of obtaining the left-hand side
of the transport equation, but one of the crucial points of this
paper that we must not use form (\ref{QKin kinet eq with coll
int}) for transport (kinetic) equation. Moreover, we should not
expect existence of collision integral in kinetic equation for
description of short range interaction of neutral particles. An
example of kinetic equation for neutral particles without
collision term is presented in Ref. \cite{Marklund EPJ B 05}.

New kinetic equation gives us another method for study of particle
dynamics. This equation is derived for a system of neutral
interacting particles. One does not contain collisions integral.
It differs from Boltzmann equation. Equation contains interaction
via term containing the trace of two-particle concentration $f_{2}$ and interaction constant. The
equation contains field variables only. In this case we can call
it field kinetic equation. We obtain equation for the case of
particles with short-range interaction. We do not use conditions
of that interaction is weak or concentration is small. We found
equation as for Bose as for Fermi particles, difference between Bose and
Fermi particles reveals at further calculations of $f_{2}$. It is interesting that we do not need to know
cross-section for particles collisions as it is in Boltzmann
equation. In this way we are free to calculate various properties
of particles system without direct knowledge inter-particle
potential of interaction. Equation (\ref{QKin kinetic equation gen SRI FOIR classic limit}) contains
information about inter-particle potential of interaction via
interaction constant. The last one is the integral of
inter-particle potential of interaction. The interaction constant can be found
from measuring of dispersion dependence
calculated from our equation. We do not need to know information
about scattering length or cross-section before solving of the
equation.

%However we need to know cross-section to solve the Boltzman
%equation.

%%%%%%%%%%%%%%%%%%%%%%%%%%%%%%%%%%%%%%%%%%%%%

\begin{acknowledgements}
The author thanks Professor L. S. Kuz'menkov for fruitful discussions.
\end{acknowledgements}

\end{document}